\documentclass[preprint,jmp,floatfix]{revtex4-1}
\usepackage{graphicx}
\usepackage{dcolumn}
\usepackage{bm}
\usepackage{amsmath}
\usepackage{amssymb}
\usepackage[utf8]{inputenc}
\usepackage[T1]{fontenc}
\usepackage{etoolbox}
\makeatletter
\def\@email#1#2{%
 \endgroup
 \patchcmd{\titleblock@produce}
  {\frontmatter@RRAPformat}
  {\frontmatter@RRAPformat{\produce@RRAP{*#1\href{mailto:#2}{#2}}}\frontmatter@RRAPformat}
  {}{}
}%
\makeatother
\begin{document}

\title{Matrix approximations of operators}

\author{B. G. Giraud$^{1\ast}$, S. Karataglidis$^{2,3\dag}$, K. Murulane$^{2\ddag}$,
  R. Peschanski$^{1\#}$}
\email{$^{\ast}$bertrand.giraud@ipht.fr}
\email{$^{\dag}$stevenka@uj.ac.za}
\email{$^{\ddag}$kmurulane@uj.ac.za}
\email{$^{\#}$robi.peschanski@ipht.fr}
\affiliation{$^1$Institut de Physique Th\'eorique, Centre d'Etudes, CEA-Saclay,
  91191 Gif-sur-Yvette, France \\
  $^2$Department of Physics, University of Johannesburg, P.O. Box 524, Auckland
  Park, 2006, South Africa\\
$^3$School of Physics, The University of Melbourne, Victoria, 3010, Australia}

\begin{abstract}
The approximate representation of operators by finite matrices is analysed
in terms of accuracy and convergence. The identity operator, for example,
can be reconstructed using a basis of harmonic oscillator states leading to a
narrow peak approximation 
of the $\delta$ function, but this peak may be perturbed by small, residual,
oscillations. The peak does
not shrink nor grows quickly, and the oscillations only diminish slowly as the
size of the matrix increases.
For the kinetic energy operator, a triple peak (one positive, two negative)
representation of $-\delta''$ is obtained,
but that is affected also by residual oscillations. Again, convergence is slow
as the matrix dimension increases.
We find compact formulas to explain such oscillations.
Similar observations are found for representations of local interactions, while
separable potentials are better represented.
As a comparison, in the context of a toy model, the effects of choosing an
alternative single particle basis are studied.  A formal approach \cite{Ka08}
for the approximation of operators  is considered for comparison.
We conclude with a word of caution for (finite) matrix approximations of
operators.
\end{abstract}
\date{\today}
\maketitle

\section{Introduction and basic formalism}
There are many instances in physics where one makes use of operators that have
simple representations in either coordinate or momentum space 
\cite{jordan2012linear,bain2006operator,bebiano2019non,hall1984kinetic}. We wish to consider representations of
such operators, whose eigenstates
may span an infinite dimensional space, as finite matrices, where that
restriction is forced by the need to diagonalise
the Hamiltonian of a system. We will consider the case of a system of $A$
identical particles whose Hamiltonian is $H = T+V$, where $T = \sum^A_{i=1} p^2_i/(2m)$
is the kinetic energy and $V = \sum_{i<j} v\left( \left| \mathbf{r}_i - \mathbf{r}_j \right| \right) + v_{cm}$
is the potential. Therein, $v$ is a two-body potential, which is often invariant under rotations and also \textit{local},
although non-local, and separable in particular, potentials have sometimes been used. The elementary mass is $m$,
while $\mathbf{p}_i \equiv \left\{ p_{xi}, p_{yi}, p_{zi} \right\}$ and $\mathbf{r}_i \equiv \left\{ x, y, z \right\}$
are the momentum and position of particle $i$, respectively. The center-of-mass (cm) potential is
$v_{\mathrm{cm}} = Am\omega^2 R^2/2$, where 
$\mathbf{R} \equiv \left\{ X, Y, Z \right\} = A^{-1} \sum^A_{i=1}  \mathbf{r}_i$
factorises the cm in a spherical
Gaussian packet at the origin of the laboratory frame, with energy
$\frac{3}{2}\hbar\omega$. This serves to remove
translational degeneracy. Without such a term, which is frequently omitted in
the literature, the ground state of
$H$ would not be square integrable as the omission leads to a zero momentum
plane wave for the system's cm.

Because $T$ and $V$ do not commute, there are very few cases where the
diagonalisation of $H$, by the solution of (integro)-differential equations,
is analytically or numerically viable.
Approximations, leading to linear or nonlinear formulations, are necessary.
The most familiar approximation is that of projecting $H$ onto a finite
dimensional subspace spanned by a set of ``basis states''
$\left| i \right\rangle$, and diagonalising the finite matrix made by the matrix elements,
$H_{ij} \equiv \left\langle i \left| H \right| j \right\rangle,\ i,j=1,...,N$,
where the basis is truncated at $N$ states.
The truncated basis may span a complete enough space but that is not usually the case.
Indeed, it has been shown that when
such a truncated basis cannot span the complete space, a convergence problem
exists \cite{Ka08,Fi25} in the search of
the solution to the Hamiltonian.
One expects that with larger $N$ generating larger embedded subspaces,
there would be convergence to the full space. The motivation of
this work is to examine such truncations with actual Hamiltonians and how a
finite set of matrix elements $H_{ij}$ might converge to
the full $H$ or not. 

The first choice to be made in this study is the choice of expansion functions. 
Given a regular local $v\left( r \right)$, the totality of suitably normalised
eigenstates $\varphi( r )$ of a one-dimensional, one-body Hamiltonian,
$h = - d^2/dr^2 + v( r )$, which may be either
discrete or continuous,  provides a resolution of the identity operator
\cite{Newt}, \textit{viz.}
\begin{equation}
  \delta\left( r - s \right) =\sum_n \varphi_n\left( r \right) \varphi^\ast_n
  \left( s \right) + 
  \int d\varepsilon \, \varphi_\varepsilon\left( r \right) \varphi^\ast_\varepsilon
  \left( s \right).
\end{equation}
This kind of expansion extends to multidimensional and multiparticle problems
by means of tensor algebra. It can also include resonant states and coupled channels
\cite{Gi04}. As a first choice, and for simplicity, we use the (one-dimensional) harmonic
oscillator basis,
\begin{align}
\varphi_1\left( r \right) & =  \pi^{-1/4} e^{-r^2/2} \nonumber \\
\varphi_2\left( r \right) & = \sqrt{2} \pi^{-1/4} e^{-r^2/2} \nonumber \\
\varphi_3\left( r \right) & = \frac{\pi^{-1/4} \left( 2r^2 - 1 \right)}{\sqrt{2}}
e^{-r^2/2} \\
& \vdots \nonumber
\end{align}
as the set of expansion functions. In Section~\ref{alt_basis}, we extend the
formalism beyond this simple basis.

In Section~\ref{identity}, we analyse how the identity operator is approximated
by finite rank kernels. A similar analysis is performed for the kinetic energy
operator in Section~\ref{kinetic} while
local and separable potentials are considered in Section~\ref{pots}. The
analyses are reconsidered in Section~\ref{alt_basis}
when switching to an alternative expansion set of states. The basis is adjusted
to a symmetry of the problem in Section~\ref{symm} and the results are 
discussed with regards to the
inclusion of the symmetry. Section~\ref{feschb} considers an alternative 
approach based on projection
operators to split the Hilbert space into two or three subspaces. We 
conclude in Section~\ref{conc}, 
comparing both approaches. An Appendix describes compact formulas 
representing matrix expansions of elementary operators onto harmonic 
oscillator wave functions.

\section{\label{identity}The identity operator in terms of the expansion basis}
We begin by calculating the kernels
\begin{equation}
  D_N\left( r, s \right) = \sum_{i=1}^N \varphi_i\left( r \right) \varphi_i
  \left( s \right).
\label{kern}
\end{equation}
The kernel in Eq.~(\ref{kern}) is invariant under exchange of $r$ and
$s$ and when both $r$ and $s$ change sign.
These kernels should approximate the function $\delta\left( r - s \right)$ and
the aim is to assess the convergence of
the function in Eq.~(\ref{kern}) to the $\delta$ function.

The expansion basis used in Eq.~(\ref{kern}) is that of a set of harmonic
oscillator states which themselves are functions
of Hermite polynomials, with the basis truncated to $N$ states. The restriction
to $N$ states provides at most
estimates of the mean square radii, $\sqrt{r^2}$ and $\sqrt{s^2}$, which
themselves are of order $\sqrt{N}$. Any results obtained
therefrom for $D_N\left( r, s \right)$ should only be considered if
$\left| r \right|$ and $\left| s \right|$ are confined to a box
of size, say $1.5\sqrt{N}$ at most.

\begin{figure}
\scalebox{0.90}{\includegraphics*{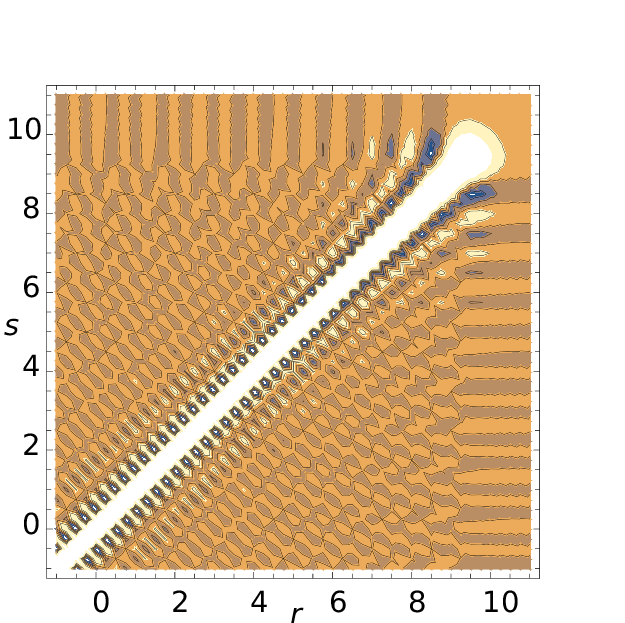}}
\caption{Contour plot of the ``approximate $\delta$-function''
  $D_N\left( r, s \right)$
obtained with $N=50$ harmonic oscillator components.}
\label{cnt50}
\end{figure}

Figure \ref{cnt50} is a contour plot of $D_{50}\left( r, s \right)$. A ridge
pattern is clearly observed along the oblique axis defined by $r=s$. As expected from
the root mean square radius argument above, this ridge strongly lowers as
soon as  $r = s \approx 7$.  It exponentially decays beyond 
$r = s \approx 10$.

\begin{figure}
\scalebox{1.00}{\includegraphics*{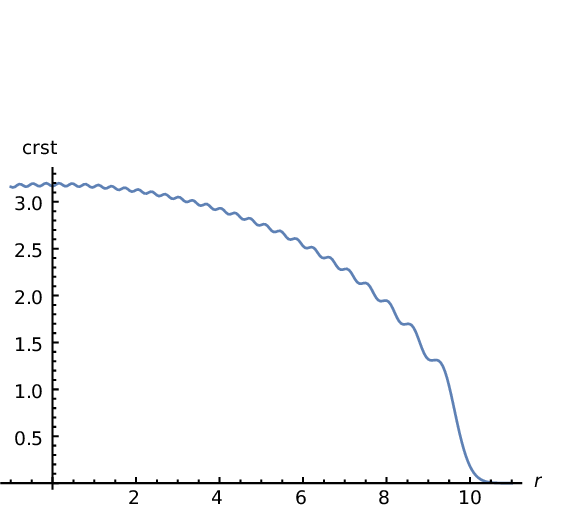}}
\caption{Crest profile $D_{50}\left( r, r \right)$ of the ridge in Fig.~\ref{cnt50}.}
\label{crst50}
\end{figure}

\begin{figure}
\scalebox{1.00}{\includegraphics*{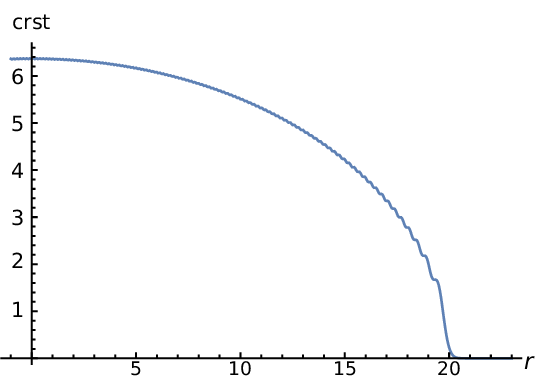}}
\caption{As for Fig.~\ref{crst50} but for $N = 200$.}
\label{crst200}
\end{figure}

Figure \ref{crst50} shows the crest profile, $D_{50}\left( r, r \right)$, of
this ridge, from which we may
make two observations. The more important one is that the profile's global
trend is of a decreasing, curved line, inducing a non negligible variation of the 
quality of the simulation of the $\delta$-function, while a plateau in some
range would have hinted a partly stable quality. When $N=200$, as shown in
Fig.~\ref{crst200}, one also does not observe such a plateau.

The second observation is the roughness of the crest. Small, but visible
oscillations are found when $N = 50$, indicating a lack of full convergence.
For $N=200$,  they become much smaller but remain.
The nature of the oscillations is clarified by the Christoffel-Darboux formula Eq.~(\ref{Apndident}),
a general property for orthogonal polynomials \cite{Eynard} (see the Appendix for generalisations),
\begin{equation}
  D_N \left( r, s \right) = \frac{1}{s-r}\sqrt{\frac{N}{2}} \left[ \varphi_{N+1}
    \left( s \right) \varphi_N\left( r \right) 
- \varphi_{N+1}\left( r \right) \varphi_N\left( s \right) \right].
\end{equation}

\begin{figure}
\scalebox{0.90}{\includegraphics*{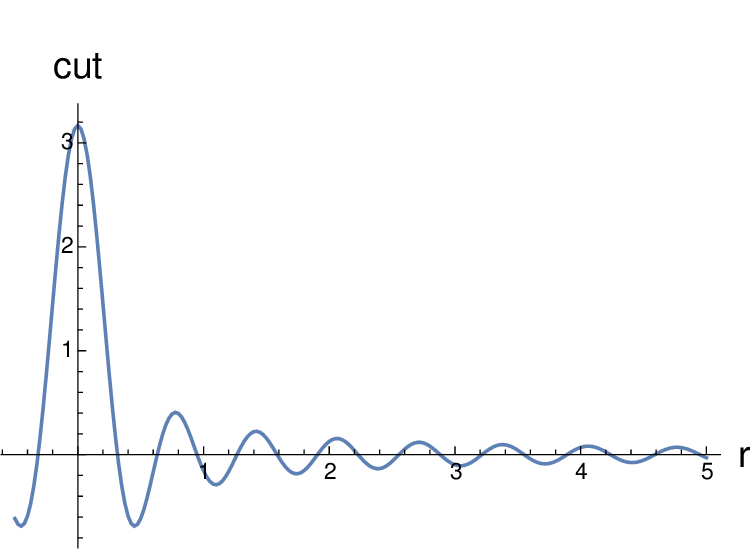}}
\caption{Shape $D_{50}\left( r, 0 \right)$ of the $s=0$ cut of the ridge when
  $N=50$.}
\label{0cut50}
\end{figure}

\begin{figure}
\scalebox{0.90}{\includegraphics*{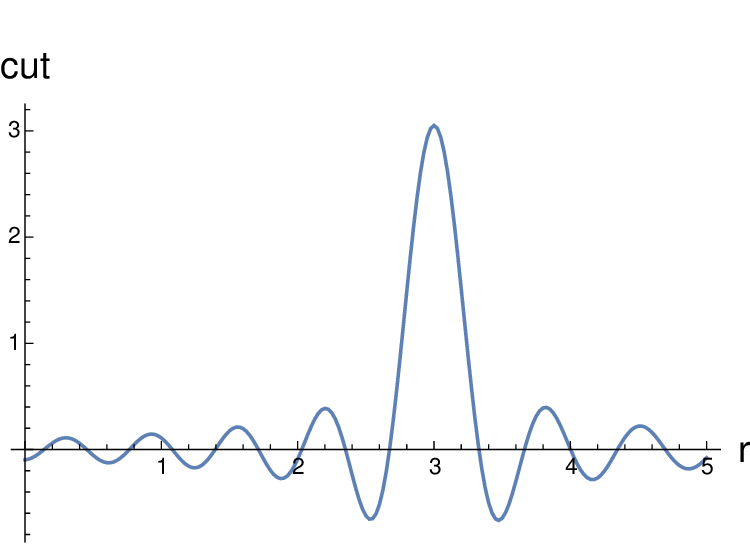}}
\caption{Shape $D_{50}\left( r, 3 \right)$ of the $s=3$ cut of the same ridge.}
\label{3cut50} 
\end{figure}

Figures \ref{0cut50} and \ref{3cut50} portray the cuts of the $N=50$ ridge at
$s=0$ and $s=3$, respectively. The shapes, $D_{50}\left( r, 0 \right)$ and
$D_{50}\left( r, 3 \right)$, appear as approximations of $\delta$ functions
centered at $r=0$ and $r=3$, respectively. They do show properly centered, somewhat high,
and somewhat narrow peaks. But they are plagued by their oscillating tails, the
influence of which may be non-negligible.

\begin{figure}
\scalebox{0.90}{\includegraphics*{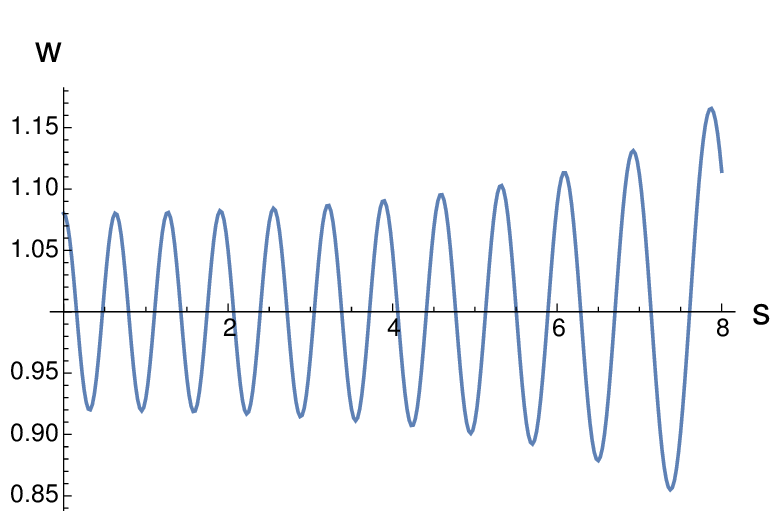}}
\caption{Oscillation of the integral $w_{50}(s)$, defined by Eq.~(\ref{cw}) with $N = 50$.}
\label{cutwei50} 
\end{figure}

\begin{figure}
\scalebox{0.90}{\includegraphics*{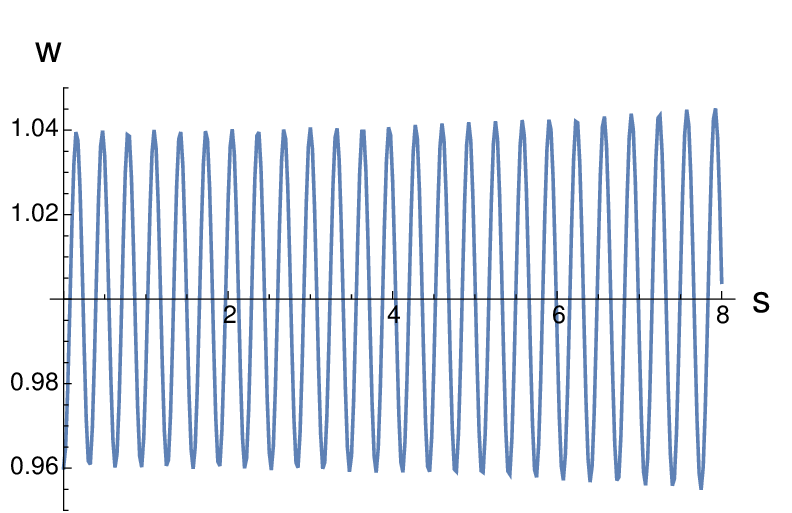}}
\caption{As for Fig.~\ref{cutwei50}, but for $N = 200$.}
\label{cutwei200}
\end{figure}

Furthermore, the ``cut weight'', defined as
\begin{equation} 
w_N\left( s \right)=\int_{-\infty}^{\infty} dr D_N\left( r, s \right), 
\label{cw}
\end{equation}
ideally should be close to $1$. Yet, as illustrated in Fig. \ref{cutwei50}, it
strongly oscillates as a function of the cut position $s$. For $N=50$ the
oscillation varies between $\approx 0.92$ and $\approx 1.08$. This amplitude
decreases very slowly as $N$ increases. For $N=200$, it still lies
between $\approx 0.96$ and $ \approx 1.04$, as shown in Fig.~\ref{cutwei200}.

The strong increase in oscillation frequency clearly comes from the growing degrees of 
the polynomials shown by Eq.~(\ref{Apndident}).

\begin{figure}
\scalebox{0.90}{\includegraphics*{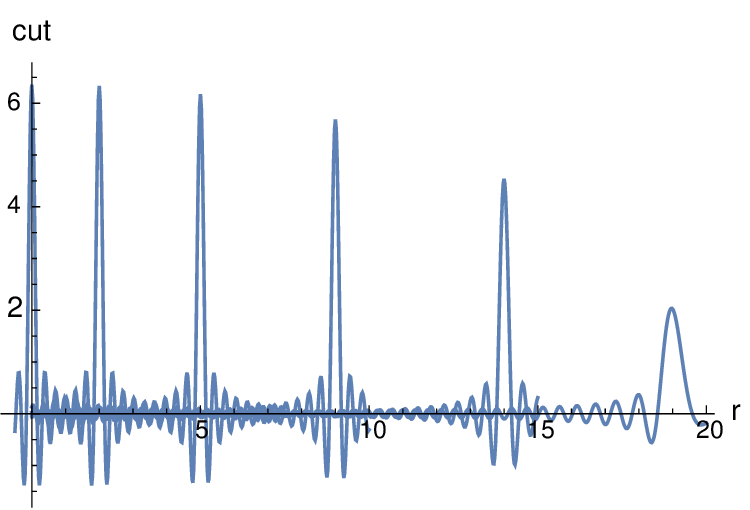}}
\caption{Cut shapes for $N=200$. Positions $s=0,2,5,9,14,19$.}
\label{cuts200}
\end{figure}

Figure \ref{cuts200} shows, for $N=200$, several cut shapes. The values of $s$
are $0,2,5,9,14$ and $19$, respectively. There is clear evolution of the shapes
between $s = 0$ and
$s = 19$, but the peaks are observed to follow the decrease of the ridge
crest. By comparison, the cut weight, $w_{200}\left( s \right)$, displayed in Fig.~\ref{cutwei200}
oscillates around the optimal value $1$.

We also produced intermediate results as a check, corresponding to $N = 100$
and $N = 150$. Those results confirm the results displayed herein, that there is some
convergence as $N$ increases, and so are not shown in the following
discussions. However, that convergence we had found to be slow.

\section{\label{kinetic}Quality of the reconstruction of the kinetic energy
  operator}
Turning our attention to the kinetic energy operator where locality features
are expected if we work in the momentum space representation, we wish to observe how this operator,
$-d^2/dr^2$, is reconstructed in the coordinate space representation. Hence, we consider kernels of the form,
\begin{equation}
\Theta_N\left( r, s \right) =
\sum_{i,j=1}^N \varphi_i\left( r \right) \left\langle i \left| \left( -
\frac{d^2}{dr^2} \right) \right| j \right\rangle 
\varphi_j\left( s \right).
\label{eqki}
\end{equation}
Again, symmetries under the transformations $\left\{r , s \right\} \rightarrow
\left\{ s, r \right\}$ and 
$\left\{ r ,s \right\} \rightarrow \left\{ -r, -s \right\}$ are trivial.

\begin{figure}
\scalebox{0.80}{\includegraphics*{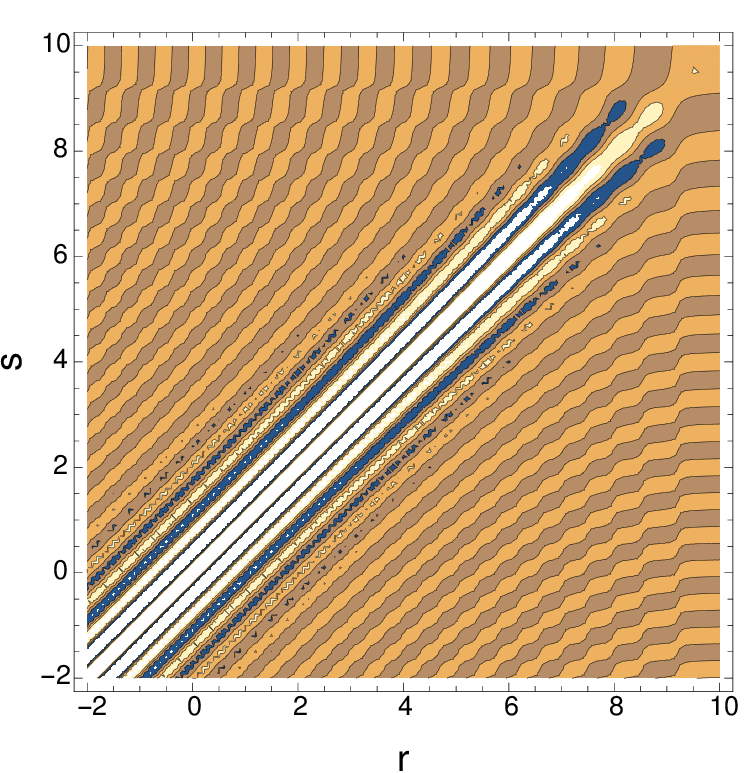}}
\caption{Reconstruction of the kinetic operator; contour plot when $N=50$.}
\label{kicnt50}
\end{figure}

\begin{figure}
\scalebox{1.0}{\includegraphics*{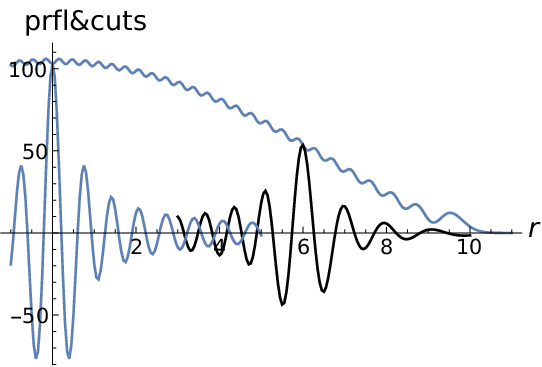}}
\caption{The ``kinetic reconstruction'' for $N=50$. The crest and cut shapes,
$\Theta(r,r)$, $\Theta(r,0)$, $\Theta(r,6)$, are displayed.}
\label{ki50cu06}
\end{figure}

Figure~\ref{kicnt50}, 
obtained with $N=50$, shows a pattern along the diagonal $r = s$. 
When a cut is implemented by freezing $s$, it exhibits a positive peak which is
flanked by two negative peaks of roughly half its surface, effectively
emulating a $-\delta''$ function. However, further oscillations do not improve matters, as
shown in Fig.~\ref{ki50cu06}, with $s = 0$ and $s = 6$ for $N = 50$. A distortion of the
shape of the cut is observed at $s = 6$, and is explained by its position near the
end of the range of validity which is limited by 50 components in the basis. The crest
ridge, as also displayed in Fig.~\ref{ki50cu06}, exhibits a shape which touches the tops of the cuts
tangentially, as
expected.

Ideally, the approximation of the second derivative of the $\delta$ function,
$\Theta_N\left( r, s \right)$,
should have a vanishing weight, given $\forall s$ by,
\begin{equation}
w_{ki}=\int_{-\infty}^\infty dr\, \Theta_N\left( r, s \right).
\label{weight}
\end{equation}

\begin{figure}
\scalebox{0.9}{\includegraphics*{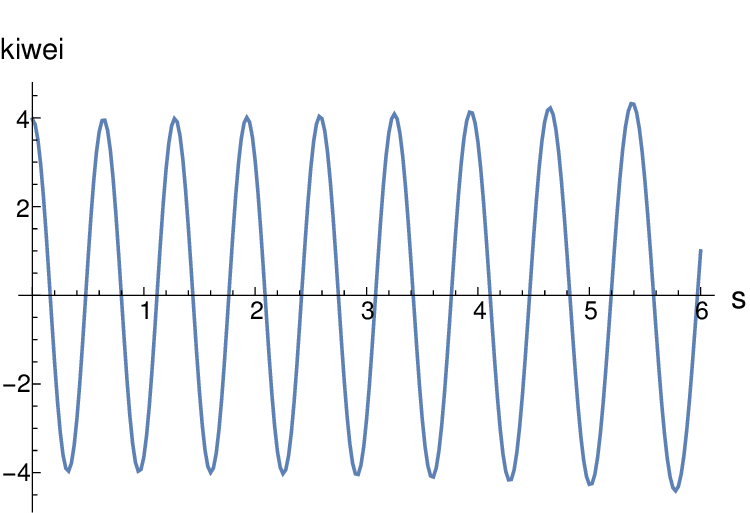}}
\caption{The weight, given by Eq.~(\ref{weight}), from the ``kinetic
  reconstruction'' of the second
derivative of the $\delta$ function for $N = 50$.}
\label{ki50badwei}
\end{figure}

However, as shown in Fig.~\ref{ki50badwei}, for $N = 50$, the weight oscillates
strongly as a function of $s$, with amplitude $\approx 4$. That amplitude grows
to $\approx 8$ for $N = 200$. This does not indicate an encouragingly fast convergence
towards the exact $\delta''$ function. The observed high oscillation frequency in the 
``noise'' of $\Theta_N$ has the same origin as that in the ``noise'' of $D_N$, namely the 
occurrence of high degree polynomials in its compact formula. See
the Appendix for the corresponding Christoffel-Darboux-like extension to that case, very
similar to Eq.~(\ref{Apndr2}).

In order to understand any convergence, we consider a different approach and
diagonalise the matrix built by the matrix elements
\begin{equation}
  \left\langle i \left| \left( - \frac{d^2}{dr^2} \right) \right| j
  \right\rangle, \ \ i,j = 1,\dots, N.
\end{equation}
From the matrix, we obtain the lowest eigenvector components
$c_i, \, i = 1, \dots, N$,
square normalised to unity, and consider the corresponding wave function
\begin{equation}
\psi_{1,N}\left( r \right) = \sum_{i = 1}^N c_i \varphi_i\left( r \right).
\end{equation}
The spectrum of the true kinetic energy operator is continuous with a lower
bound of 0. Hence, in the present matrix approximation, the wave function,
$\psi_{1,N}\left( r \right)$, should be as smooth and flat as possible to simulate a wave without momentum.
Indeed, the waves $\psi_{1,50}$ and $\psi_{1,200}$ are positive parity and reasonably
flat. The former extends between $\approx \pm 10$ until it exponentially decays. It
reaches a maximum of 0.32. For the latter, the range and maximum values read $\approx
\pm 20$ and $\approx 0.22$, respectively. Hence, at the cost of multiplying the number
of components by 4, one obtains ratios of $\approx 2$ for the ranges, which seems reasonable,
and $\approx 0.22/0.43 = 0.69$ for the flatness. A rule of thumb for square normalisation as the range
doubles would predict a scaling of $1/\sqrt{2} \approx 0.71$ for the wave function, which is close to the
observed ratio. These latter considerations may attenuate the possible doubts suggested by the 
oscillations seen above.

\section{\label{pots}Reconstruction of potentials}
%\label{pots}
\subsection{Simple, local potentials}

A local potential, $v$, by definition, is given by the matrix element between
states $\left| r \right\rangle$ and
$\left| s \right\rangle$, \textit{viz.}
\begin{align}
  \left\langle r \left| v \right| s \right\rangle & = v\left( r \right)
  \delta\left( r - s \right) \nonumber \\
& = v\left( \frac{r+s}{2} \right) \delta\left( r - s \right),
\end{align}
where the second equation is of a more symmetric form. With this definition of
the local potential, we may form the kernels
\begin{equation}
  \mathcal{V}_N\left( r, s \right) = \sum_{i,j = 1}^N \varphi_i\left( r \right)
  \left\langle i \left|
 v \right| j \right\rangle \varphi_j\left( s \right)
\label{eqpot}
\end{equation}
and assess if the kernel forms a reasonable representation of the local
potential. Further, we may investigate whether the intuitive ansatz, $D_N\left( r, s \right)
v\left( \frac{r+s}{2} \right)$ itself makes
a reasonable approximation of $\mathcal{V}_N$ if $D_N$ is a suitable
approximation of the $\delta$ function
as discussed in Section~\ref{identity}.

\begin{figure}
\scalebox{0.90}{\includegraphics*{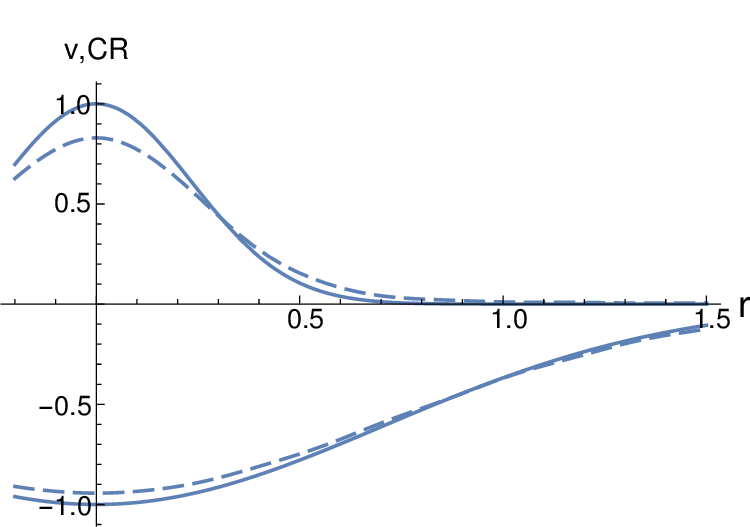}}
\caption{The potentials $v$ (solid lines) along the crest of their
  finite matrix kernels $\mathcal{V}$ for $N=50$. The crest ratios
  [Eq. (\ref{cr})] are portrayed by
the dashed lines. (See also the discussion in text.)}
\label{potcret}
\end{figure}

We first consider the local potentials $v\left( r \right) = - e^{-r^2}$ and
$v\left( r \right) = e^{-9r^2}$.
The (one-dimensional) harmonic oscillator basis used herein has as its ``germ''
the basic Gaussian $\pi^{-1/4}\beta^{-1/2} e^{-r^2/(2\beta^2)}$, with the length 
scale $\beta = 1$. The two 
potentials then can be considered as a moderate-range attractive potential and
a short-range repulsive potential, respectively. The crest ratio (hereafter CR), defined as
\begin{equation}
CR = \frac{\mathcal{V}_N\left( r, r \right)}{D_N\left( r, r \right)},
\label{cr}
\end{equation}
is displayed in Fig.~\ref{potcret}, for $N = 50$.

Therein, the crest ratios are portrayed by the dashed lines while the
potentials themselves are
portrayed by the solid lines. The approximation of the moderate-range potential
shows an error of at most $\approx 6\%$ when comparing the potential to its CR
while the
error in the approximation for the short-range potential worsens to
$\approx 17\%$. When one
increases $N$ to a value of 200, the errors reduce to $3\%$ and $8\%$,
respectively, with the
reduction in the error of a factor of 2, indicating a reasonable convergence
to the actual potentials.

\subsection{The case of $r^2$}

The reconstruction of the kinetic energy operator was somewhat problematic.
Similar problems may be expected for the case of reconstruction of $r^2$ since, except for pure
phases under a Fourier transformation, harmonic oscillator wave functions are the same in
both the coordinate and momentum space representations. The reconstruction of the $r^2$
operator in coordinate space is a perfect image of the reconstruction of its kinetic
partner, $p^2$, in momentum space. We define
\begin{equation}
\mathcal{R}_N\left( r, s \right) = \sum_{i,j = 1}^N \varphi_i\left( r \right)
\left\langle i \left| r^2 \right| j \right\rangle \varphi_j\left( s \right).
\label{eqr2}
\end{equation}
A compact form of this sum is available in the Appendix, see Eq.~(\ref{Apndr2}).
Figure~\ref{cnt50r2} shows the contour plot of $\mathcal{R}_{50}\left( r, s
\right)$. There is a notable ridge pattern along the diagonal axis, $r = s$. If one compares the
\begin{figure}
\scalebox{0.80}{\includegraphics*{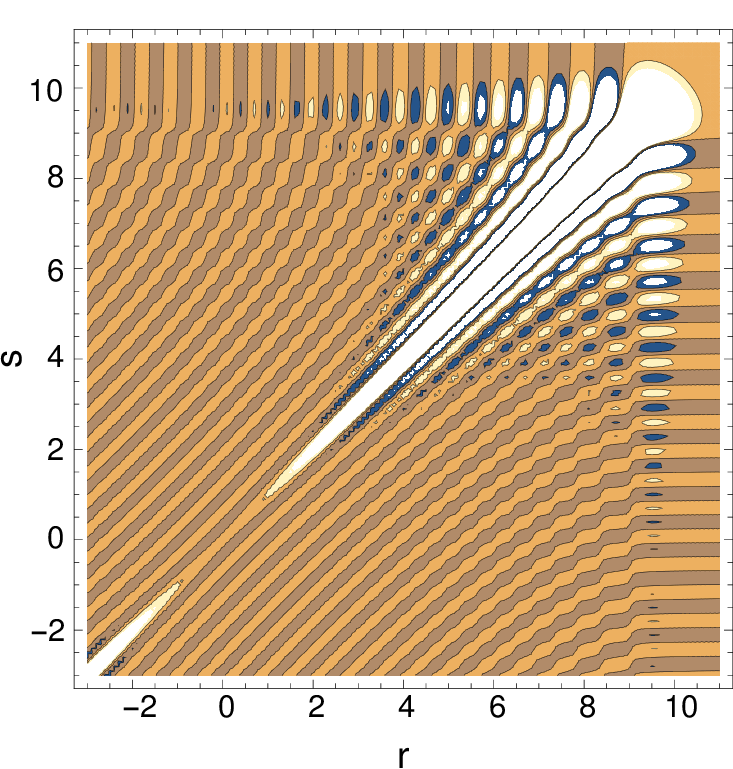}}
\caption{The contour plot of $\mathcal{R}_{50}\left( r, s \right)$, the
  reconstruction of the $r^2$
operator.}
\label{cnt50r2}
\end{figure}
results in Fig.~\ref{cnt50r2} with those of Fig.~\ref{cnt50}, one finds more oscillations in the contour plot in
Fig.~\ref{cnt50r2}, a low region around the origin, $r = s = 0$, and then a rise in the ridge
until an exponential collapse when either $\left| r \right|$ or $\left| s \right|
\approx 11$. Recall, that in Section~\ref{identity} for the case of $N = 50$, the range beyond which the
basis could not be trusted was $\approx 7$. See again the Appendix for the rich pattern of
oscillations.

\begin{figure}
\scalebox{0.80}{\includegraphics*{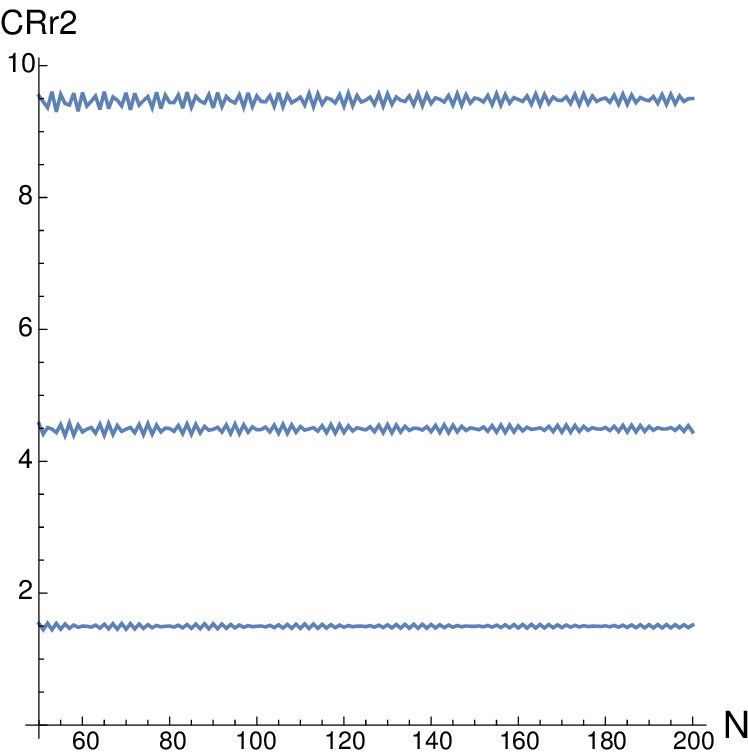}}
\caption{The reconstruction of $r^2$. The horizontal lines are crest ratios as
  functions of $N$. 
$r=1, 2, 3$, the CR fluctuates about $1.5,4.5,9.5$, respectively.}
\label{CRr2}
\end{figure}

At the origin, $r = s = 0$, the CR is worth $\frac{1}{2}$ and does not depend
on $N$. The ratios, $\mathcal{R}_N\left( 1, 1 \right)/D_N\left( 1, 1 \right)$,
$\mathcal{R}_N\left( 2, 2 \right)/D_N\left(2,2\right)$, and
$\mathcal{R}_N\left( 3, 3 \right)/D_N\left( 3, 3 \right)$, are
shown in Fig.~\ref{CRr2} as functions of $N$. They fluctuate around $1.5$, $4.5$ and
$9.5$, respectively. More generally, the ratio $\mathcal{R}_N\left( r, r \right)/D_N\left( r, r \right)$ converges
slowly towards $r^2 + 1/2$ as $N$ increases.

By evaluating the eigenstates of the matrix
\begin{equation}
  \rho_N = \left\{ \left\langle i \left| r^2 \right| j \right\rangle, \,
  i, j = 1, \dots, N
\right\},
\end{equation}
these observations of an insufficient quality of the matrix reconstruction of
$r^2$
may be completed. We let $N$ be the number of basis components and $c_j$ the
components, square normalised to unity, of any of the eigenvectors of $\rho_N$,
with $\lambda$ the corresponding eigenvalue. Hence
\begin{equation}
\sum_j \rho_{kj} c_j = \lambda\, c_k.
\end{equation}
If we consider the wave function,
\begin{equation}
\psi\left( r \right) = \sum_j c_j \varphi_j\left( r \right)
\end{equation}
we find that $\psi$ is an eigenstate of the kernel $\mathcal{R}$ given that
\begin{equation}
\int^\infty_{-\infty} \mathcal{R}\left( r, s \right) \psi\left( s \right) \, ds = 
\lambda \psi\left( r \right).
\end{equation}
And, given that $\mathcal{R}$ may be considered an acceptable approximation of
$r^2$, then $\psi$ may be considered a reasonable approximation of an
eigenstate of the operator $r^2$.

\begin{figure}
\scalebox{1.00}{\includegraphics*{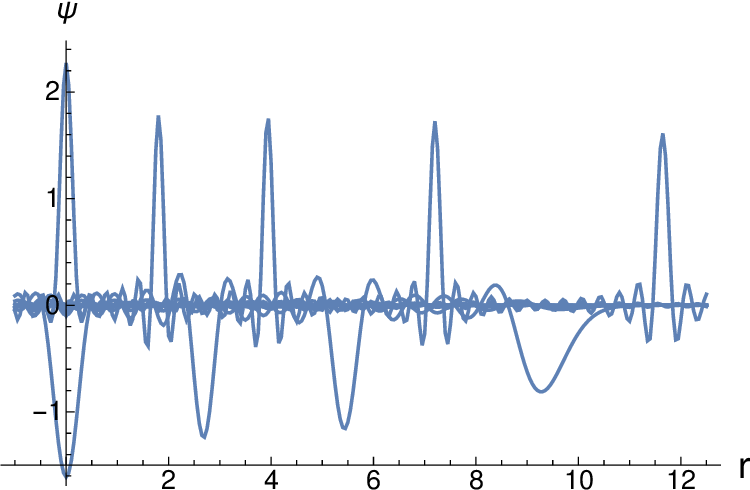}}
\caption{The following localised states, as obtained by
  diagonalising finite matrix approximations of $r^2$, are displayed. Peaks
  above the ${\bf r}$-axis:
$\psi_{1,200}$, $\psi_{30,200}$, $\psi_{50,200}$, and $\psi_{120,200}$.
Below the axis: $\psi_{1,50}$, $\psi_{30,50}$, and $\psi_{50,50}$. (See
the text for details.)}
\label{pic50200r2}
\end{figure}

For the following, we return to the original definitions of the variables:
the index $i$
is again associated with the eigenstate of $\rho_N$, represented by the
$\left\{ c_j \right\}$
and where $N$ is the component number. As $r^2$ is a local operator whose eigenstates
are represented by $\delta$ functions, we expect that the $\psi_{i,N}$ become
narrower as $N$ increases. That is the case as shown in Fig.~\ref{pic50200r2}.
Note that a slight, somewhat irrelevant, complication is brought by parity, a 
property of the matrix $\rho_N$. Except for the ground states $\psi_{1,N}$, which
are centered at the origin, the good parity states $\psi_{i,N}, \, i = 2, \dots,
N$ are made
of pairs of peaks. For graphical convenience, Fig.~\ref{pic50200r2}, shows only
those peaks that are on the positive side of the $r$ axis.

The eigenvalues for the ground states, in the cases of $N=50$ and $N=200$, are
$\lambda_{1,50} = 0.0244$ and $\lambda_{1,200} = 0.0062$. The width of the wave
packet $\psi_{1,200}$ is close to half that of $\psi_{1,50}$, and its height is
multiplied
by $\approx \sqrt{2}$. This is expected and an indication of convergence as $N$
increases.
Along with the ground states, Fig.~\ref{pic50200r2} shows $\psi_{30,50}$,
$\psi_{50,50}$,
$\psi_{30,200}$, $\psi_{50,200}$, and $\psi_{120,200}$. For graphical convenience,
and
for comparison purposes, the peaks for $N=200$ are shown as mainly positive
functions while
those peaks for $N=50$ are negative functions. It is evident that the peaks
decrease in
width and increase in height as $N$ increases. This confirms the convergence,
albeit a
slow one.

One may also consider whether the sum $\Theta + \mathcal{R}$ [(Eqs.~(\ref{eqki})
  and (\ref{eqr2})], which in these cases gives exact eigenvalues,
$\Lambda_i = \left( 2i - 1 \right)$, while separately with each generated term giving somewhat imprecise
results,
would still yield reasonable results if the choice of the functions
$\left\{ \varphi_i \right\}$
were different. Difficulties with Eqs.~(\ref{eqki}) and (\ref{eqr2}) are 
almost 
guaranteed, as these attempt to reproduce operators that have continuous 
spectra
by operators of finite rank. But, with a different basis, the sum
$\Theta + \mathcal{R}$,
being still an approximation of $-d/dr^2 + r^2$, and likely having a discrete
spectrum,
may still permit good approximations of eigenvalues, given that errors at first
order in wave functions translate to second order for the eigenvalues. See 
Section~\ref{alt_basis} for a numerical illustration.

\subsection{Separable potentials}

Let $\left| \xi \right\rangle$ denote the form factor, assumed to be real, so
that the
potential becomes separable: $\xi(r) \xi(s)$. Eq.~(\ref{eqpot}) may then be
written as
\begin{equation}
  \mathcal{W}_N\left( r, s \right) = \left[ \sum_{i=1}^N \varphi_i
    \left( r \right) \left. \left\langle
    i \right| \xi \right\rangle \right] \left[ \sum_{j=1}^N \left. \left\langle
    \xi \right| j \right\rangle \varphi_j \left(
s \right) \right].
\label{eqsep}
\end{equation}
The quality of the separable representation is governed by the quality of the
form factor,
$\left| \xi_N \right\rangle = \sum^N_{i = 1} \left. \left\langle i \right| \xi
\right\rangle
\left| i \right\rangle$. The question of convergence reduces to that of convergence in
the context of square integrability.

\subsection{Semi-realistic potential}

\begin{figure}
\scalebox{0.90}{\includegraphics*{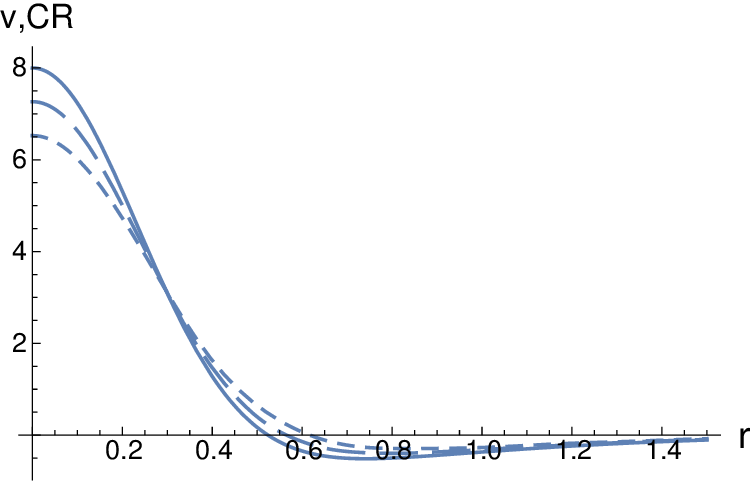}}
\caption{The reconstructed semi-realistic potential using the CR method. See
  text for details.}
\label{reastic}
\end{figure}

We use the potential 
\begin{equation}
v\left( r \right) = 9 e^{-9r^2} - e^{-r^2},
\label{srealv}
\end{equation}
which combines a strong repulsive core and a mid-range attraction. Figure
\ref{reastic} shows the potential as a full line, its somewhat acceptable
reconstruction by the ``CR method'' with $N=200$ as a long dash
curve, and the weaker result for $N=50$ as a smaller dash curve. Overall, the
results of the approximations using the CR method are acceptable, except near $r = 0$.

\subsection{Extension to a toy Hamiltonian} 

In this case we continue the investigation by generalising to a toy Hamiltonian
\begin{equation}
h = -\frac{d^2}{dr^2} + e^{-9r^2} - e^{-r^2}
\label{toyham}
\end{equation}
where we have mixed the repulsive core and the attractive term in proportions
that allow for a bound state, simplifying the numerics. The exact ground state,
$\psi_{\mathrm{ex}}$, has the eigenvalue $\varepsilon_{\mathrm{ex}} = - 0.1720763$.
We then project $h$ onto subspaces with bases of, for example, 50 and 100
of the oscillator wave functions. (The subspaces may be further reduced to
bases of 25 and 50 states due to parity conservation.) The ground state
eigenvalues found
in those subspaces are $\varepsilon_{50} = - 0.171874$ and $\varepsilon_{100} =
-0.172071$, respectively. This is a reasonable result and the comparison of the
respective ground states shows quite small errors of the order of $10^{-2}$ and
$10^{-4}$ for $N = 50$ and $N = 100$, respectively.

\section{\label{alt_basis}Different single particle basis}
%\label{alt_basis}
The simplicity and convenience of the harmonic oscillator basis does not always
guarantee fast enough convergence. We now consider an alternative set of basis functions,
namely an even number $N$ of translated Gaussians,
\begin{equation}
\varphi_i\left( r \right) = \pi^{-\frac{1}{4}} \beta^{-\frac{1}{2}} 
\exp\left[ - \frac{\left( r + \frac{N+1}{2} \sigma - i\sigma \right)^2}
  {2\beta^2} \right],
i = 1, \dots, N.
\label{shiftgauss}
\end{equation}  
Therein, $\sigma$ is a shift parameter.
The basis formed by these functions exhibits, except for edge effects, an
approximate translational invariance. However, the basis needs to be orthonormalised into a
set of functions $\left\{ \chi_i \left( r \right), i = 1,\dots, N \right\}$. The
orthonormalisation process is not unique and we tested two processes; the
results do not depend on the process, as expected. 

When $\sigma$ is significantly smaller than the width of the Gaussians, taken
again with $\beta = 1$, the orthogonalisation process involves derivatives and can
reconstruct something similar to the harmonic oscillator basis. For that reason,
we take $\sigma = \beta$ instead. This choice separates each Gaussian from
its neighbours, somewhat moderately, leading to finite difference effects rather
than approximate derivatives. It also defines an efficiency range, of the
order of $N\sigma/2$ at the minimum. Upon inspection, one of our new bases
may still show states with some similarity to the oscillator states but the
corresponding width parameter, $\beta_{\mathrm{new}}$, would be significantly
larger than $\beta$.

When $N=50$, the first seven lowest eigenvalues of the matrix approximation
of $-d^2/dr^2+r^2$ with this new basis are,
$\{1.00043,3.00005,5.05263,7.0184,9.6656,11.3889,15.7908 ...\}$, to be
compared with the exact sequence $\{1,3,5,7,9,11,13 ...\}$. While the
approximation
is seen to be quite good at first, an onset of worsening accuracy
occurs after the $5^{\mathrm{th}}$ or $6^{\mathrm{th}}$ eigenvalue. This is to
be expected.

We return to the approximation of the $\delta$ function which, in 
terms of the normalised new basis states, is written as
$\Delta_N\left( r, s \right)
= \sum^N_{i=1} \chi_i\left( r \right) \chi_i\left( s \right)$. And, to confirm
our numerical results, we also generated the same, self-replicating, kernel as
$\Delta_N\left( r, s \right) = \sum^N_{i=1} \varphi_i\left( r \right)
\tau_i\left( s \right)$, introducing the dual space spanned by the basis
states $\tau_i$ with $\left. \left\langle \tau_i \right| \varphi_j \right\rangle = \delta_{ij}$;
that basis can be obtained by inverting the matrix of scalar products $\left. \left\langle
\varphi_i \right| \varphi_j \right\rangle$. For $N = 50$ and the new
basis of shifted Gaussians, the contour plot of $\Delta_{50}$ resembles that
displayed in Fig.~\ref{cnt50} for $D_{50}$. Yet significant differences exist.

\begin{figure}
\scalebox{0.90}{\includegraphics*{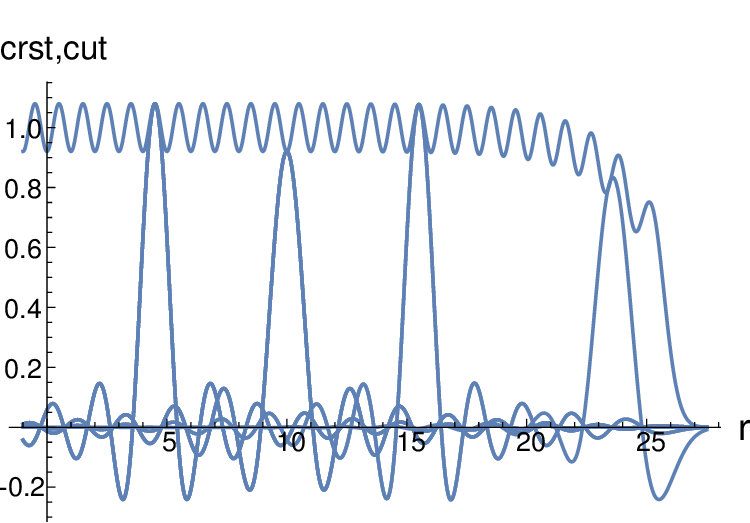}}
\caption{The profile of the crest, and of the cuts at $s = 4.5, 10, 15.5,
  23.5$, using
the new basis with $\beta=\sigma=1,\; N=50$.}
\label{nbcrstcut50}
\end{figure}

\begin{figure}
\scalebox{0.90}{\includegraphics*{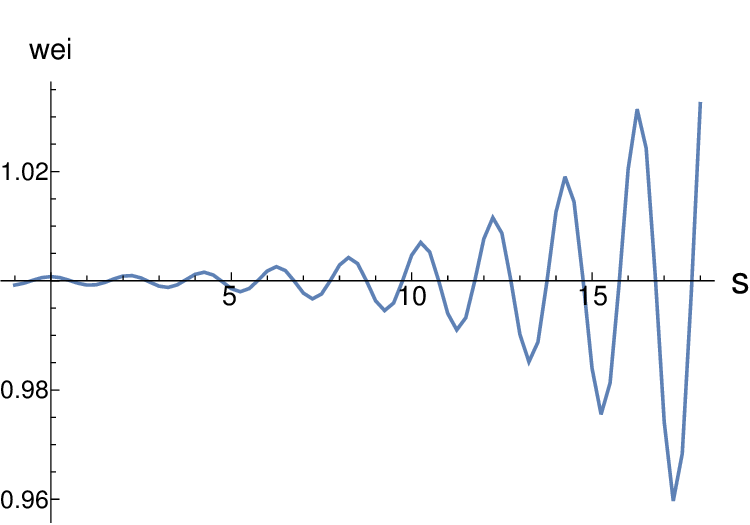}}
\caption{The weight function, Eq.~(\ref{dweight}), obtained using the
same basis as that in Fig.~\ref{nbcrstcut50}.}
\label{nbmas50}
\end{figure}

As shown in Fig.~\ref{nbcrstcut50}
the crest now shows an almost horizontal trend except for residual oscillations.
Also, with an ``efficiency range'' of $\approx 12$, the cuts seems to be similar
to one another, suggesting a reasonable stability in the approximation. Further,
as long as $\left| s \right|$ is less than $\approx 12$, the ``weight'',
defined by
\begin{equation}
w = \int_{-\infty}^{\infty} \Delta\left( r, s \right) \, dr,
\label{dweight}
\end{equation}
remains close to 1, as illustrated in Fig.~\ref{nbmas50}.

\begin{figure}
\scalebox{1.2}{\includegraphics*{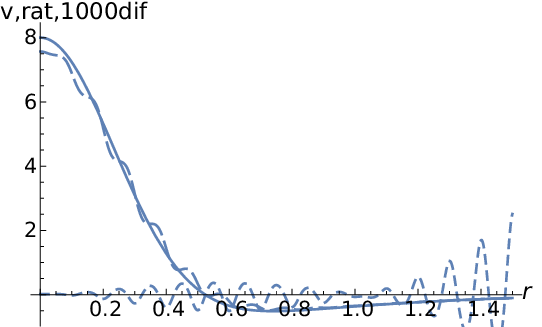}}
\caption{The potential. The solid line portrays the full function. The
  long-dashed  line is that which is approximated using a basis of 60 shifted Gaussians,
  using the CR  method. The short-dashed line is the expanded difference between the potential
  created using 60 basis states and that created using 40 basis states, \textit{viz.},
  $1000\left( v_{60} - v_{40} \right)$.}
\label{pot4060}
\end{figure}

As a final study with an alternative basis, we now reconstruct the
semi-realistic potential given by
Eq.~(\ref{srealv}). However, as the potential has an exponential
decay beyond $r \approx 2$, it is not very useful to consider shifted Gaussians
beyond this value of $r$. We choose instead a basis set,
\begin{equation}
\varphi_i\left( r \right) = \pi^{\frac{1}{4}} \beta^{-\frac{1}{2}} \exp\left[
- \frac{\left( r - r_c \right)^2}{2\beta^2} \right],
\label{secgauss}
\end{equation}
where $r_c = i\sigma - \left( N + 1/2 \right)\sigma, \; i = 1,\dots, N$. We
take $\beta = \sigma
= 0.1$, for an example, and set $N = 40$ and $N=60$ as two comparison
reconstructions.
The former value of $N$ provides an effective range of $\approx 2$ and the
latter value a
range of $\approx 3$. Figure~\ref{pot4060} shows the results of the
reconstructions, where $v_{40}$ and $v_{60}$ are the potentials found for $N = 40$ and $60$,
respectively, both using the CR method. In that figure, the actual results for both values of
$N$ are practically equal, and so only $v_{60}$ is portrayed. Indeed, both give
very reasonable reproductions of the original potential. To highlight the
difference between the two results, we also give the difference $v_{60} - v_{40}$ multiplied by 1000.

\section{\label{symm}Single particle basis taking advantage of symmetry}
%\label{symm}
We consider again the toy Hamiltonian of Eq.~(\ref{toyham}). We wish to
compare its exact ground state, $\psi_{\mathrm{ex}}$, with eigenvalue $\varepsilon_{\mathrm{ex}} =
-0.172076$, with the approximate ground state $\psi_N$ and its eigenvalue when 
the Hamiltonian is projected onto one of the subspaces. The ground state is even
under parity and so we consider a ``raw'' basis of pairs of basis states
\begin{equation}
  \varphi_i\left( r \right) = \frac{1}{\sqrt{2}}\pi^{\frac{1}{4}} \beta^{-\frac{1}{2}}
  \left[ 
e^{-\left( r - r_c \right)^2/\left( 2\beta^2 \right)} + e^{-\left( r + r_c \right)^2/
\left( 2\beta^2 \right)} \right],
\label{symbas}
\end{equation}
where $r_c = \left( i - \frac{1}{2} \right) \sigma, \; i = 1,\dots, N$, and
with the
understanding that this basis is constructed before orthonormalisation.

\begin{figure}
\scalebox{0.90}{\includegraphics*{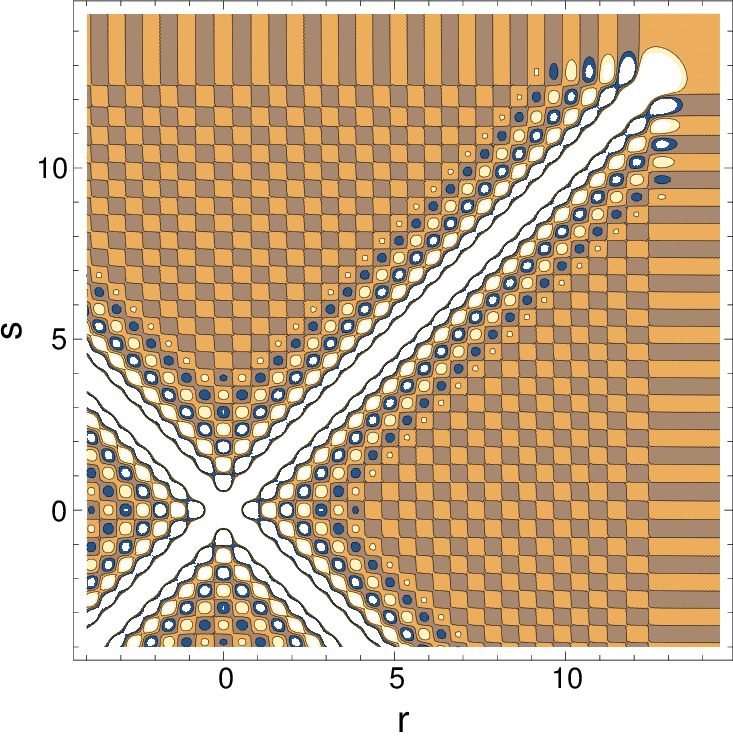}}
\caption{The contour plot of $\Delta$ found using a basis of 25 pairs
of symmetrically shifted Gaussians with $\beta = \sigma = \frac{1}{2}$.}
\label{cntplu25}
\end{figure}

Figure~\ref{cntplu25} shows the contour plot of the idempotent kernel,
$\Delta_N\left( r, s \right) = \sum^N_{i=1} \chi_i\left( r \right)
\chi_i\left( s \right)$,
for $N = 25$ and $\beta = \sigma = 1/2$. This choice spreads the basis states,
and their orthonormalised forms $\chi_i$, in a range $\left| r \right|
\lesssim 12$.
Note, that for the line $r = -s$, there is an additional ridge due to parity.
Small interference effects are noticeable near $r = s = 0$, which are due to
those few wave packets of the raw basis that are close to the origin. The pair
of ridges provide an approximation of the sum
$\left[ \delta\left( r - s \right) + 
\delta \left( r + s \right) \right]/2$, from which there has to be
a positive interference near the origin. That is observed in Fig.~\ref{cntplu25}
when $r,s \lesssim 1$. The crest profile shows a flat trend,
perturbed by the expected oscillations, between $r = s \approx 1$, and
$r = s \approx 10$.
This suggests the stability of the approximation in those domains.

\begin{figure}
\scalebox{1.2}{\includegraphics*{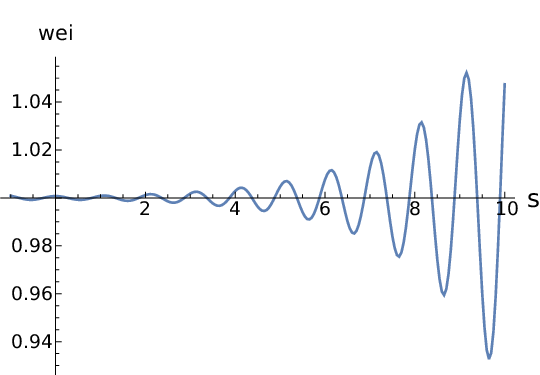}}
\caption{The basis from 25 pairs of Gaussians. Also displayed is the evolution
  of the weight $w(s)$
of a cut at given $s$.}
\label{mssplu} 
\end{figure}

For $N = 25$, the approximate ground state of the toy Hamiltonian has
an eigenvalue $\varepsilon_{25} = -0.17194$. compared to the exact value
of $\varepsilon_{\mathrm{ex}} = -0.17208$. The eigenvalue for $N = 25$ is
a reasonable result, better than the value obtained using harmonic oscillator
states at $N = 50$, $\varepsilon_{50} = -0.17187$. The ``weight'',
$w\left( s \right) = \int_{-\infty}^{\infty} \Delta\left( r, s \right) \, dr$
remains
close to $1$, with limited oscillations, in the range $0.96 \lesssim
w\left( s \right)
< 1.04$. The observed range of $w\left( s \right)$ is within a somewhat large
domain $\left| s \right| \lesssim 8$, as shown in
Fig.~\ref{mssplu}, despite its fragmentation between the simulations of
$\frac{1}{2}\delta\left( r - s \right)$ and
$\frac{1}{2}\delta\left( r + s \right)$. Finally, the difference between the
approximate wave function and
the exact one for this toy Hamiltonian, $\psi_{25} - \psi_{\mathrm{ex}}$, is of
order $\approx 10^{-2}$ at most.

In the case of $N = 50$, an accuracy of one more decimal place is required
given the way
$\varepsilon_{50}$ converges to the exact eigenvalue, and also in the way
$\psi_{50}$
converges to the exact wave function. Two more decimal places are found to
see any variation
in $w\left( s \right)$ near unity. Those good results are partly misleading,
however, given that
we took the precaution of verifying the reconstructions of the kinetic and
potential energy
operators in earlier sections. For such operators individually no truly
satisfactory result was obtained; looking at
the Hamiltonian in full produced a far better convergence.

\section{Effective forces}
\label{feschb}
We now consider the Feshbach projection formalism, namely
to split the full Hilbert space into two or three subspaces. (See refs
\cite{Fe58,Fe62}.)
Following the approach of Karataglidis and Amos \cite{Ka08} (hereafter KA),
the subspaces are created
by the use of projectors $\mathcal{P} = \sum_{i=1}^{N_1} \left| \chi_i
\right\rangle \left\langle \chi_i
\right|$, $\mathcal{Q} = \sum_{i=N_1+1}^{N_2} \left| \chi_i \right\rangle
\left\langle \chi_i
\right|$ and $\mathcal{R} = 1 - \mathcal{P} - \mathcal{Q}$. The subspaces are
formed by
projecting out the solutions to the Hamiltonian using these projectors.
However, it may come to pass
that the subspaces $P$ and $Q$ are not enough to span the entire Hilbert
space and so the third
subspace is required. The splitting of the Hilbert space into subspaces is
necessary when truncation 
is required in the Hamiltonian matrix to allow for diagonalisation: the $P$
subspace (the
``included'' space) is formed by
the states which are found through direct diagonalisation of the truncated
Hamiltonian, while
the $Q$ subspace (the ``excluded'' space) are those states that are not found
through
the diagonalisation. The effective Hamiltonian, which is reconstructed by use
of the projection
operators and the kernel, contains coupling between the $P$ and $Q$ subspaces,
and $R$ if
necessary, to converge to the exact solutions.

\begin{figure}
\scalebox{0.90}{\includegraphics*{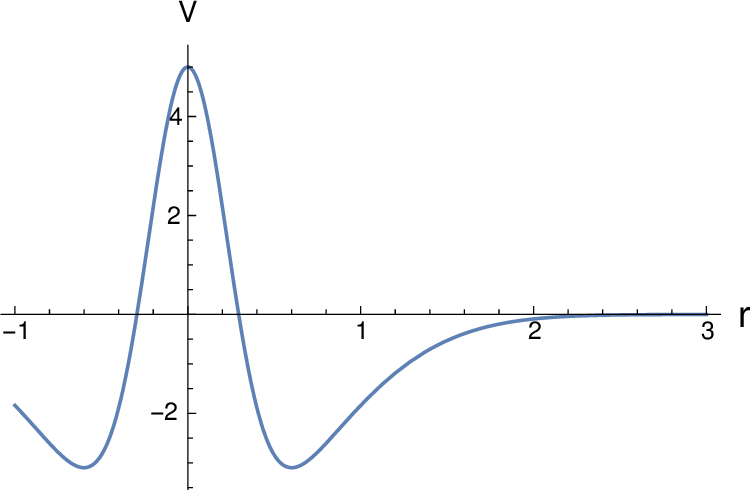}}
\caption{Toy potential for a Feshbach method analysis.}
\label{wpot} 
\end{figure}

For a one-dimensional toy model, for which we choose the Hamiltonian
\begin{equation}
h = -\frac{d^2}{dr^2} + 10e^{-9r^2} - 5e^{-r^2},
\label{hfeschb}
\end{equation}
we use a basis of harmonic oscillator even-parity states,
and choose $N_1 = 5$ and $N_2 = 26$. This allows for a small $P$ space with the
$Q$ space providing a significant correction to any results in the $P$ space.
The potential in Eq.~(\ref{hfeschb}) is shown in Fig.~\ref{wpot}.

\begin{figure}
\scalebox{0.90}{\includegraphics*{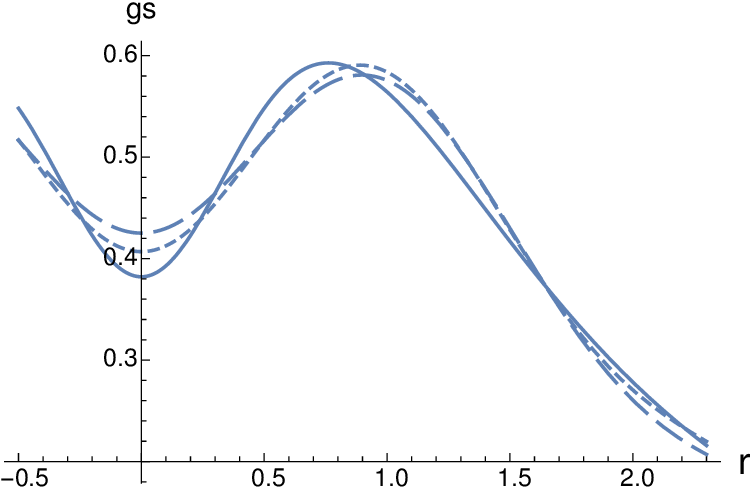}}
\caption{Ground states for the Hamiltonian, Eq.~(\ref{hfeschb}).
The results of the various calculations are given by: the solid line for
$\left( \mathcal{P} + \mathcal{Q} \right) h \left( \mathcal{P} + \mathcal{Q}
\right)$; 
the long dash one for $\mathcal{P} h \mathcal{P} + W_{\mathrm{eff}}\left( E_0
\right)$; 
and the small dash one for $\mathcal{P} h \mathcal{P}$. }
\label{wigwaves} 
\end{figure}

Consider first the $P$ and $Q$
subspaces only. The required matrices are $\mathcal{P}h\mathcal{P}$,
$\mathcal{P}h{Q}$,
$\mathcal{Q}h\mathcal{P}$, $\mathcal{Q}h\mathcal{Q}$, with the kernel
$G\left( E \right) =
\left( E - \mathcal{Q}h\mathcal{Q} \right)^{-1}$, from which we define the
effective kernel $W_{\mathrm{eff}}
\equiv \mathcal{P}h\mathcal{Q}G\left(E \right) \mathcal{Q}h\mathcal{P}$. The
diagonalisation of
$\left( \mathcal{P} + \mathcal{Q}\right) h \left( \mathcal{P} + \mathcal{Q}
\right)$ yields a ground state
energy $E_0 = -0.7342$ which compares favorably with the ground state
eigenvalue, $-0.7342256$, obtained
by numerical solution of the corresponding differential equation. (Notice, 
incidentally, that this availability of a precise numerical solution means that
we have now activated the $R$ subspace). The same diagonalisation of
$\left( \mathcal{P} + \mathcal{Q}\right) h \left( \mathcal{P} + \mathcal{Q}
\right)$ also gives a very good
approximation for the ground state wave function, $\psi_0$. The comparison
with the approximate
reconstructions are shown in Fig.~\ref{wigwaves}.
Therein, the solid line is the best approximation. By construction, the
diagonalisation of $\mathcal{P}h
\mathcal{P} + W_{\mathrm{eff}}$ gives the same energy eigenvalue
$E_0 = -0.7342$ and the
projected ground state wave function $\mathcal{P}\psi_0$. The projected ground
state,
renormalised, is shown by the long-dashed line, and differs from that found by 
diagonalising only $\mathcal{P} h \mathcal{P}$, the result of which is portrayed
by the short-dashed line. That wave function has a corresponding eigenvalue of
$E_0^{\mathrm{sub}} = -0.6897$, and which predictably under-binds.

The results obtained from such a toy model are sufficient to vindicate a use of
matrix projections
for \textit{well-chosen} subspaces as tools for understanding and correcting
eigenvalues and
eigenstates, although the previous sections show the limits of such
projections. This
cautious yet positive appraisal comes from the availability of powerful
programs for numerical
diagonalisations and inversions of matrices, even those with large dimensions. 

A further note of caution. There is a convergence problem when
$\mathcal{P}$ and $\mathcal{Q}$ are defined in such a way that, however large is their size,
their is no way in which they might span the full Hilbert space. In that case, the
$R$ subspace, courtesy
of its projector $\mathcal{R}$, restores the spanning of the Hilbert space.
And if the $R$
space is required, then convergence to the exact solution of both eigenvalues
and states is difficult as the number of intermediate couplings in the effective Hamiltonian
between the $Q$ and $R$ subspaces may be infinite \cite{Ka08}.

\begin{figure}
\scalebox{0.90}{\includegraphics*{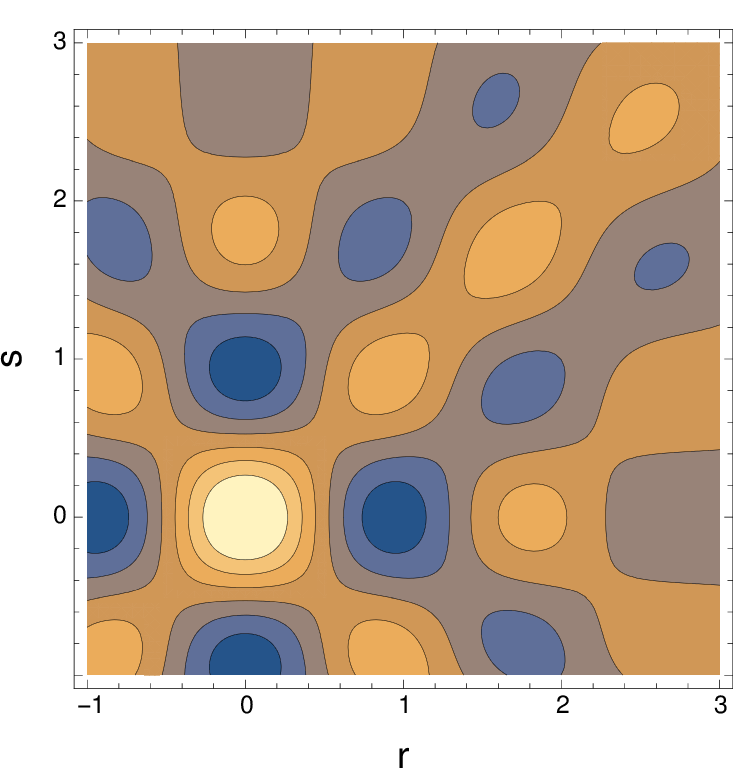}}
\caption{$N=5$ contour plot for $\mathcal{P}h\mathcal{P}$. There is a vague hint
of quasi locality along the diagonals $r = \pm s$.}
\label{phpfig}
\end{figure}

\begin{figure}
\scalebox{0.90}{\includegraphics*{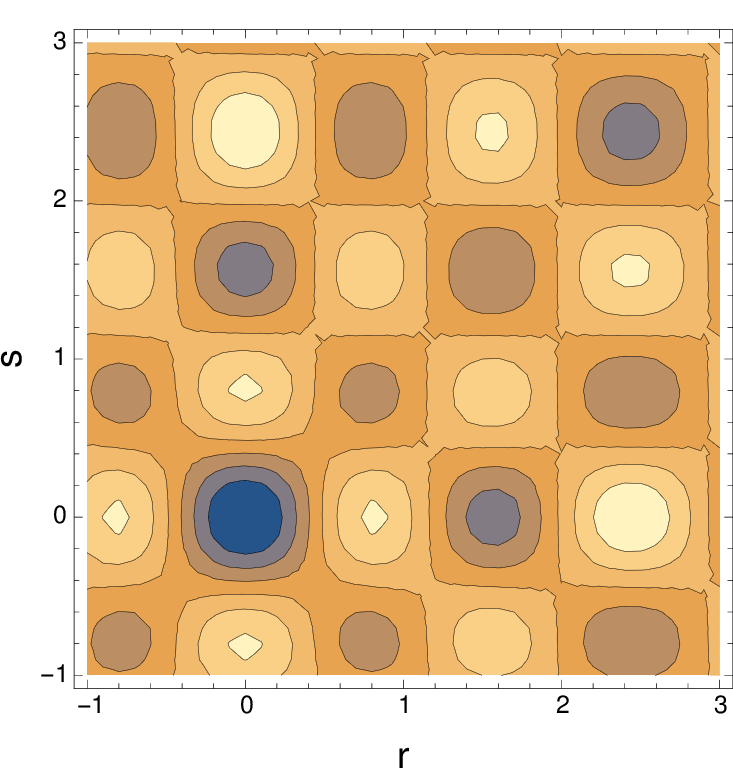}}
\caption{As for Fig.~\ref{phpfig}, but for the effective potential
  $W_{\mathrm{eff}}$.}
\label{weffpot}
\end{figure}

This leads to the question of the effective potential: can $W_{\mathrm{eff}}$ be
used to define a convenient, additional, energy-dependent, and (more or less)
local
potential for a diagonalisation in the $P$ subspace? Formally, given the
highlighted convergence problem, the answer would be negative. We show in
Figs.~\ref{phpfig} and \ref{weffpot} the contour
plots for $\mathcal{P}h\mathcal{P}$ and $W_{\mathrm{eff}}$, respectively. The
contour plot
for $\mathcal{P}h\mathcal{P}$ shows approximate ridge patterns,
located around the
lines $r = \pm s$, as expected for the quasi-locality of the toy Hamiltonian.
However, the
effective potential shown in Fig.~\ref{weffpot} has no indication of any
locality; the non-locality
in $W_{\mathrm{eff}}$ is extreme.
 
\section{Summary and Conclusions}
\label{conc}
Coined by Heisenberg, the term ``matrix mechanics'' extended the selection and 
description of atomic, molecular, and nuclear states from the bare quantum numbers 
of shells to a broader class of configuration mixings. The resulting calculations, ranging 
from valence mixings to large shell-model configuration mixings, are extensive. The list of 
references \cite{prakash1969configuration, mcgrory1967shell, mcgrory1980large, 
draayer1984symplectic, brown2006new, teran2006behavior, grawe2004shell, 
caurier2005shell, coraggio2009shell, kato2011brueckner, Ka19} represents only a tiny 
fraction of the full literature. In particular, the case of hard-core interactions, requiring 
resummations over infinite subsets of configurations to obtain effective interactions, defines 
a specific domain related to configuration mixings \cite{nesbet1958brueckner, de1994n, 
haxton2001canonical, stroberg2019}.

The standard view in matrix mechanics is that, since operators can be represented 
by matrices, one must balance between small matrices that are practical and larger ones 
that offer improved approximations at greater computational cost. It is also generally assumed 
that increasing the matrix size yields better convergence. 

Most matrix representations employ a discrete and finite orthonormal basis of configurations. 
(We are aware of the generator coordinate method \cite{griffin1957collective, wong1975generator, 
reinhard1987generator, verriere2020time,Gi73}, but it was not used here due to complications 
introduced by its norm kernel.) The essential point is that discrete orthogonal bases necessarily 
produce oscillating wave functions, which in turn generate interference effects in the matrix elements 
of operators and in resummations. These effects slow down the rate of convergence of the approximations. 

We thus studied how elementary operators such as the identity, kinetic energy, harmonic 
potential, and toy potentials are represented in different bases. The analysis also included 
the Feshbach approach formulated in terms of complementary subspaces. 

In the case where a suitable finite basis was chosen to truncate the Hilbert 
space to finite dimension, calculations using simple one-dimensional harmonic 
oscillators showed that convergence toward the exact operators 
(in the case of the $\delta$ function), or wave functions
and their eigenvalues, is possible, but one cannot escape that convergence
cannot be fast. Yet, for the same finite number of basis states, a different
expansion set of basis functions
may yield better convergence. Such a choice may indeed be critical in cases
where practical calculations are required.

When a splitting of the Hilbert space into two or three subspaces is necessary
the issue is more serious. 
KA~\cite{Ka08} identified a convergence problem when three subspaces, $P$,
$Q$, and $R$ are required and when only two of the three, $P$ and $Q$, are
used in calculations.
The neglect of the third subspace may lead to serious convergences problems. In
our case, where
we investigated the diagonalisation of a toy Hamiltonian in the truncation of
the included space,
there is still a degree of locality, but the results indicated the non-locality
that is introduced is
due to the neglect of the third subspace. Further, the effective interaction
becomes highly
non-local, whereas the potential used in the toy Hamiltonian is clearly local.

One must take care in either scenario when taking results of calculations
which are
approximations to the exact physical problem. The question remains as to which
approximation(s)
are better in such usage. All told, we find that matrix approximations can
be mathematically valid, but caution must be exercised concerning the choice 
of the approximation subspace
and the rate of convergence.

\appendix
\section{Generalisations of the Christoffel-Darboux formula}
\label{appendix}
We know that the idempotent kernel,
\begin{equation}
D_N\left( r,s \right)=\sum_{i=1}^N \varphi_i(r) \varphi_i(s)
\end{equation}
can be reduced to a simple, compact form, 
\begin{equation}
  D_{\mathrm{cmpct}}\left( r, s \right) = \alpha(N) {\left( s - r \right)}^{-1}
  \left[
    \varphi_{N+1}\left( s \right) \varphi_N\left( r \right)-
    \varphi_{N+1}\left( r \right)
    \varphi_N\left( s \right)) \right],
\label{Apndident}
\end{equation}
 where $\alpha(N)=\sqrt{\frac{N}{2}}$ for the harmonic oscillator basis. See
 \cite{Eynard} for the general case of orthogonal polynomials.
 
A similarly simpler form is available for the kernel
\begin{equation}
  \mathcal{R}_N\left( r, s \right)  = \sum_{i,j=1}^N
  \varphi_i\left( r \right) \langle i \left| r^2 \right| j \rangle
  \varphi_j\left( s \right).
\end{equation}
From this we may write
\begin{align}
  r^2 D_N & = \sum_{i=1}^N r^2 \varphi_i\left( r \right) \varphi_i\left( s \right)
  \nonumber \\
& = \sum_{i=1}^N \sum_{j=1}^{N+2}
  \varphi_j\left( r \right) \left\langle j \left| r^2 \right| i \right\rangle
  \varphi_i \left( s \right).
\end{align}
This is equivalent to
\begin{multline}
r^2 D_N = \mathcal{R}_N\left( r, s \right)  \\ 
+ \varphi_{N+1}\left( r \right) \frac{\sqrt{N(N+1)}}{2}
\varphi_{N-1}\left( s \right) + 
\varphi_{N+2}\left( r \right) \frac{\sqrt{(N+1)(N+2)}}{2}
\varphi_{N}\left( s \right),
\label{Apndr2}
\end{multline}
upon taking advantage of the symmetry and selection rules of the
matrix element $\left\langle j \left| r^2 \right| i \right\rangle $. This
provides $\mathcal{R}_N(r,s)$
in terms of $r^2$ times the compact form $D_{\mathrm{cmpct}}$ of $D_N$, and of a
pair of corrective terms.

It is trivial to extend the present formalism to polynomials in $r$ and
$d/dr$. For instance, given $h=r^2-d^2/dr^2$, the projected Hamiltonian
$D_N h D_N$ benefits from a compact form,
\begin{equation}
  D_N h D_N=r^2 D_{\mathrm{cmpct}}-D''_{\mathrm{cmpct}},
\end{equation}
where the double prime denotes the second derivative with respect to $r$.
No additional term is needed here, because the $\varphi_i$'s are eigenstates
of $h$.

We recall here that such compact forms are of interest for the understanding
of the spurious oscillations often introduced by truncations.

\bibliography{local_sk}

\end{document}